\documentclass{aa}

\usepackage{epsfig}
\usepackage{graphicx}

\def\gtaprx {\lower .1ex\hbox{\rlap{\raise .6ex\hbox{\hskip .3ex
 {\ifmmode{\scriptscriptstyle >}\else {$\scriptscriptstyle >$}\fi}}}
 \kern -.4ex{\ifmmode{\scriptscriptstyle \sim}\else
 {$\scriptscriptstyle\sim$}\fi}}} 
\def\ltaprx {\lower .1ex\hbox{\rlap{\raise .6ex\hbox{\hskip .3ex
 {\ifmmode{\scriptscriptstyle <}\else {$\scriptscriptstyle <$}\fi}}}
 \kern -.4ex{\ifmmode{\scriptscriptstyle \sim}\else
 {$\scriptscriptstyle\sim$}\fi}}} 
\def\etal {et al. }
\def\littleprime{\ifmmode{\scriptscriptstyle \prime }
 \else{\hbox{$\scriptscriptstyle \prime$ }}\fi}
\def\littless{\ifmmode{\scriptscriptstyle s }
 \else{\hbox{$\scriptscriptstyle s $ }}\fi}
\def\littlemm{\ifmmode{\scriptscriptstyle m }
 \else{\hbox{$\scriptscriptstyle m $ }}\fi}
\def\littlehh{\ifmmode{\scriptscriptstyle h }
 \else{\hbox{$\scriptscriptstyle h $ }}\fi}
\def\littlecirc{\ifmmode{\scriptscriptstyle \circ }
 \else{\hbox{$\scriptscriptstyle \circ $ }}\fi} 
\def\rasec{\raise .9ex \hbox{\littless}} 
\def\arcsec{\raise .9ex \hbox{\littleprime\hskip-3pt\littleprime\hskip-3pt}} 
\def\ramin{\raise .9ex \hbox{\littlemm}} 
\def\arcmin{\raise .9ex \hbox{\littleprime}}
\def\hrs{\raise .9ex \hbox{\littlehh}} 
\def\deg{\hbox{$^\circ$}}
\def\degree{\raise .9ex \hbox{\littlecirc}} 
\def\magpoint{\hbox to 2pt{}\rlap{\hskip -.5ex \arcmm}.\hbox to 2pt{}} 
\def\arcsspoint{\hbox to 1pt{}\rlap{\arcss}.\hbox to 2pt{}} 
\def\arcsecpoint{\hbox to 1pt{}\rlap{\arcsec}.\hbox to 2pt{}} 
\def\arcminpoint{\hbox to 1pt{}\rlap{\arcmin}.\hbox to 2pt{}} 
\def\degreepoint{\hbox to 1pt{}\rlap{\degree}.\hbox to 2pt{}}
\def\nodata{ ~$\cdots$~ }
\def\lax{{$\mathrel{\hbox{\rlap{\hbox{\lower4pt\hbox{$\sim$}}}\hbox{$<$}}}$}}
\def\gax{{$\mathrel{\hbox{\rlap{\hbox{\lower4pt\hbox{$\sim$}}}\hbox{$>$}}}$}}

\begin{document}

\title{Light-year Scale Radio Cores in Four LINER Galaxies}

\author{Mercedes E. Filho\inst{1}, Peter D. Barthel\inst{1}, \and 
Luis C. Ho\inst{2}}
\offprints{mercedes@astro.rug.nl}

\institute{Kapteyn Astronomical Institute, P.O.~Box 800,
NL--9700\,AV Groningen, The Netherlands \and
The Observatories of the Carnegie Institution of Washington,
813 Santa Barbara St., Pasadena CA 91101, USA}

\date{Received date; accepted date}

\titlerunning{Radio Cores in LINERs}
\authorrunning{Filho, Barthel, and Ho}

\abstract{The  LINER  galaxies  NGC\,2911, NGC\,3079,  NGC\,3998,  and
NGC\,6500 were observed  at 5~GHz with the European  VLBI Network at a
resolution  of 5~milliarcsecond  and found  to  possess flat-spectrum,
variable, high-brightness temperature  ($T_{\rm B}\,>\, 10^8$~K) radio
cores.   These radio  characteristics  reinforce the  view that  these
LINERs   host  central   engines  associated   with   active  galactic
nuclei.   }

\maketitle

\section{Introduction}

As many as  40\% of all nearby galaxies display  some level of nuclear
activity  qualitatively  resembling  that  seen  in  accretion-powered
active  nuclei (Ho  \etal 1997b;  Ho 1999a).   The  activity manifests
itself in  objects such as  Seyfert nuclei and  low-ionization nuclear
emission-line  regions  (LINERs; Heckman  1980).   LINERs differ  from
Seyferts   in   that    they   display   characteristically   stronger
low-ionization optical forbidden lines.   The source of the ionization
responsible for the optical emission  in LINERs is still under debate.
Models which  invoke shock heating (Fosbury \etal  1978; Heckman 1980;
Dopita  \&  Sutherland 1995),  stellar  photoionization (Terlevich  \&
Melnick 1985;  Shields 1992;  Filippenko \& Terlevich  1992; Terlevich
\etal   1992;   Barth  \&   Shields   2000),   and  aging   starbursts
(Alonso-Herrero \etal 2000) have  been proposed as possible ionization
mechanisms.  There  is, however,  growing evidence that  a substantial
fraction  of   the  LINER  population  simply   constitute  the  local
low-luminosity  equivalent  of  ``classical'' active  galactic  nuclei
(AGN) such as quasars and  luminous Seyfert galaxies (see review by Ho
2001).   This  is borne  out  by the  detection  of  weak radio  cores
(Heckman 1980; Sadler \etal 1989;  Wrobel \& Heeschen 1991; Slee \etal
1994;  Nagar \etal  2000; Falcke  \etal 2000),  by the  nature  of the
ultraviolet  (Maoz  \etal 1995,  1998;  Barth  \etal  1998) and  X-ray
radiation  (Ptak \etal  1999;  Terashima \etal  2000; Halderson  \etal
2001; Ho \etal 2001), and by  the presence of broad H$\alpha$ lines in
total (Ho  \etal 1997c) as well  as polarized flux  (Barth \etal Moran
1999).   If  many  LINERs are  in  fact  true  AGNs, this  would  have
repercussions on  the faint  end of the  AGN luminosity  function, and
consequently on galaxy evolution and the cosmic X-ray background.

Here we report  on an investigation of the  radio morphologies of four
LINER galaxies on milliarcsecond  (mas) scales.  Observations with the
European  VLBI  Network  (EVN)  were  used to  obtain  images  of  the
light-year scale  structure in  the radio cores  of these  four LINERs
suspected to be powered by AGNs.

This paper  adopts a Hubble constant  of $H_{\rm 0}$  = 75 km~s$^{-1}$
Mpc$^{-1}$  and defines the  spectral index,  $\alpha$, such  that the
flux density $F_{\nu}\, \propto\, \nu^{-\alpha}$.

\section{Sample Selection}

The four  LINER galaxies  studied here (Table~1)  were taken  from the
Palomar spectroscopic survey  of bright, northern galaxies (Filippenko
\& Sargent 1985;  Ho \etal 1995, 1997a, b, c).   They were selected as
having  fairly bright radio  emission, with  significant contributions
from  an  unresolved component  on  arcsecond  (VLA) scales.   Optical
images  of NGC\,2911,  NGC\,3079  and  NGC\,3998 can  be  seen in  the
Sandage \& Bedke (1994) {\it Carnegie Atlas of Galaxies}, and an image
of  NGC\,6500 is available  in Gonz\'alez  Delgado \&  P\'erez (1996).
Various  lines  of  evidence,  to  be discussed  below,  suggest  that
accretion-driven  power plays  a  role  in the  nuclei  of these  four
galaxies.

Early VLBI  experiments conducted on  these sources have  yielded some
correlated flux  densities on M$\lambda$ baselines  (van Breugel \etal
1981;  Jones   \etal  1981;  Hummel  \etal   1982).   However,  images
permitting  assessment  of  morphological  properties  and  brightness
temperatures  had  not  been  obtained.  We  therefore  conducted  EVN
observations to obtain such images and to quantify the contribution of
the radio emission arising from  the AGN and the starburst components.
In addition, it was hoped that mas-scale structure could be traced out
to large (kpc) scales.  We note, in passing, that a typical EVN 5\,GHz
resolving  beam of  5 mas  translates to  1.5--3 light  years  for the
sources under study.

%TABLE 1 - Summary

\begin{table*}[!ht]
\footnotesize
\begin{center}  
\parbox[b]{17.0cm}{\vspace{2mm}
\caption{{\bf Target Galaxies}. 
Column 1: Source name. 
Column 2: Spectral class of the nucleus from Ho et al. 1997a, where L = LINER, S 
= Seyfert, 1.9 = weak broad H$\alpha$ emission line present, and 2 = no broad 
emission line; NGC 3079 falls on the borderline between Seyferts and LINERs.  
Column 3 and 4: Optical position from NED (NASA/IPAC Extragalactic Database).
Column 5: Adopted distance from Tully 1988, who also uses our value of 
$H_{\rm 0}$.
Column 6: Hubble type from NED.
Column 7: Green Bank 1.4\,GHz flux density from White \& Becker 1992, 
12\arcmin~resolution.
Column 8: NVSS 1.4\,GHz flux density from Condon et al. 1998, 
45\arcsec~resolution. 
Column 9: FIRST 1.4\,GHz flux density from Becker,  White, \&  Helfand 1995, 
5\arcsec~resolution. 
Column 10: Green Bank 4.9\,GHz flux density from  Becker,  White, \&  Edwards
1991, 3\arcminpoint5 resolution. 
}
}
\begin{tabular}{cccccccccc}
\hline
Source & Spectral & R.A.(J2000)            & Dec.(J2000) & $D$   & Hubble & GB\,1.4 & NVSS\,1.4 & FIRST\,1.4 & GB\,4.9 \\

       & Class    & (\hrs \ramin \rasec) & (\degree \arcmin \arcsec) & (Mpc) & Type   & (mJy)   & (mJy)     & (mJy)      & (mJy) \\

(1)    & (2)      & (3)                    & (4) & (5)   & (6)    & (7)     & (8)       & (9)        & (10) \\

\hline 

NGC\,2911 & L2    & 09 33 46.10\  & 10 09 08.5\ & 42.2 & S0:  pec & \nodata & \ 58.6 & \nodata & \ 73 \\
NGC\,3079 & S2/L2 & 10 01 57.30\  & 55 40 54.0\ & 20.4 & SBc      & 845     & 770.7 & 293.0 & 321 \\
NGC\,3998 & L1.9  & 11 57 56.112  & 55 27 12.74 & 21.6 & S0?      & 126     & 101.4 & \ 98.5 & \ 82 \\
NGC\,6500 & L2    & 17 55 59.771  & 18 20 18.32 & 39.7 &  Sab:    & 224     & 182.9 & \nodata & 176 \\ 

\hline

\end{tabular}
\end{center}
\end{table*}
\normalsize

\section{Observations and Data Reduction}

The  observations of NGC\,3998  and NGC\,6500  were performed  on 1997
June 5--6, while NGC\,2911 and NGC\,3079 were observed on 1997 June 9,
all with the EVN Mark\,III  system at 5\,GHz.  We combined 14 channels
of 4\,MHz each to achieve a total bandwidth of 56\,MHz.  The requested
antennas were  Effelsberg, Jodrell Bank Mk\,2,  Medicina, Noto, Torun,
Westerbork, and Onsala, but  unfortunately only the first five yielded
useful   data.    Total   integration   time,   aiming   for   optimal
$uv$-coverage,  was  about   4~hours  per  galaxy.   Cross-correlation
employed the Bonn (MPIfR) Mark\,III correlator.

All of  the Onsala data  were lost, and  the data from  the Westerbork
array were also lost due  to wrong polarization observation.  The loss
of these  two antennas  was equivalent to  losing 47$\%$ of  our data.
Furthermore, more than half of  the 14 Torun channels were lost during
the  observations   of  NGC\,3998  and   NGC\,6500.   Following  cross
correlation,  subsequent  data  reduction  was performed  at  JIVE  in
Dwingeloo.   After minor  flagging, the  data were  calibrated, fringe
fitted,  and Fourier  transformed using  standard tools  in  AIPS (van
Moorsel  \etal   1996).   Typical  angular   resolution  achieved  was
5--8\,mas (Gaussian FWHM).  The  uncertainty of the absolute amplitude
calibration, mainly due to data noise and uncertainties in the primary
flux density calibrator, is estimated to be 5\%--10\% (1\,$\sigma$).

\section{Results}

All four  LINERs were detected at  5\,GHz and found  to display strong
($\sim$20--80 mJy) point sources.  Table~2 lists the properties of the
maps and  the main measured  quantities, while Table~3  summarizes the
derived quantities.  Weak  extended emission, at a few  percent of the
peak level,  is observed in NGC\,3079, NGC\,3998,  and NGC\,6500; this
will be discussed below.  The deconvolved images are not shown because
they suffer from sidelobe remnants  and noise peaks.  Because our maps
are dominated by a highly compact central core, and because we are not
confident about the robustness of the faint extended emission, we have
chosen  not to  perform  detailed fits  to  obtain deconvolved  source
sizes.

Peak brightness temperatures were computed as 

\begin{center}
$$T_{\rm B} = \frac{F_{\rm peak} \, c^2} {2 \, k_{\rm B} \, 
\nu^2 \, \Omega^2} \hspace{0.5cm} {\rm (K),} $$
\end{center}

\noindent where $F_{\rm peak}$ is the peak 5\,GHz flux density, $c$ is
the speed of light, $k_{\rm  B}$ is Boltzmann's constant, $\nu$ is the
observing frequency, and  $\Omega^2$ is the upper limit  to the source
size. We conservatively regard all the cores to be unresolved and give
upper limits to  their sizes, equivalent to half  of the Gaussian FWHM
of the  synthesized beam. Given that  the resolution effects  are at a
few percent level  at most, the combined use of  peak flux density and
the  adopted   $\Omega^2$  yields  lower  limits   to  the  brightness
temperature figures.

The  next  section  discusses  the  EVN  imaging  results  within  the
framework of other properties known for these galaxies.

%TABLE 2 - Measured Radio Parameters

\begin{table*}[!ht]
\footnotesize
\begin{center}
\parbox[b]{10.0cm}{\vspace{2mm}
\caption{{\bf Measured Radio Parameters}.
Column 1: Source name.
Column 2: Beam size (FWHM).
Column 3: Position angle of the beam.
Column 4: Rms noise of the image.
Column 5: Peak 5~GHz flux density.
Column 6: Integrated 5~GHz flux density.
}
}
\begin{tabular}{cccccc}
\hline
Source    & Beam Size           & PA    & rms               & $F_{\rm peak}$ & $F_{\rm int}$             \\
          & (mas $\times$ mas ) & (\deg)& (mJy beam$^{-1}$) & (mJy)          & (mJy) \\
(1)       & (2)                 & (3)   & (4)               & (5)            & (6)                        \\
\hline
NGC\,2911 &\ 8.4 $\times$ 5.7   & 50.4  & 0.19              & 18.7           & 19.9                   \\
NGC\,3079 &\ 7.2 $\times$ 5.1   & 58.5  & 0.20              & 13.8           & 14.8                   \\
NGC\,3998 &\ 7.4 $\times$ 4.7   & 25.9  & 0.28              & 78.2           & 83.0                    \\
NGC\,6500 & 10.1 $\times$ 3.6   & 42.1  & 0.24              & 68.5           & 83.9                     \\
\hline
\end{tabular}
\end{center}
\end{table*}
\normalsize

\section{Individual Galaxies}

\subsection{NGC\,2911} 

The  dominant radio  core  in NGC\,2911  shows pronounced  variability
(Jones \etal  1982; Wrobel \&  Heeschen 1984; Condon \etal  1991; Slee
\etal 1994).  From arcminute- and arcsecond-resolution observations it
appears that about 10~mJy  of low-surface brightness radio emission is
present along the  major axis of the galaxy  (Wrobel \& Heeschen 1984;
Condon \etal 1991).  Weak mas structure at PA $\approx$ $-$30\deg\ was
claimed to  have been  detected in an  early 1.7\,GHz  VLBI experiment
(Jones  \etal   1981),  but  was   not  confirmed  with   5\,GHz  VLBI
observations (Jones \etal 1982; Schilizzi \etal 1983).

We do  not detect  mas-scale extended emission.   Our results  show an
unresolved,  high-brightness temperature ($T_{\rm  B} >  10^8$~K) core
with   a   flux    density   of   $\sim$20~mJy.    Long-baseline   PTI
(Parkes-Tindinbilla interferometer) measurements  in the late 1980s by
Slee \etal  (1994) yielded  a relatively flat  radio spectrum  for the
core; $\alpha$ = 0.21 between  2.3 and 8.4~GHz.  Interpolating the PTI
flux densities to 5\,GHz yields  a value of 43~mJy, which, compared to
the 163~mJy  measured in 1980 with  the VLA (Wrobel  \& Heeschen 1984)
and to our  1997 result, confirms the presence  of strong variability.
The presence of a variable, unresolved, flat-spectrum, high-brightness
temperature radio core constitutes compelling evidence for an AGN-type
source in this LINER galaxy.

\subsection{NGC\,3079} 

Arcsecond-resolution  images  of this  well-known  active galaxy  show
large-scale radio structure. Emission  along the galactic disk and two
15-kpc nonthermal lobes emanate from  the nucleus along the minor axis
(de Bruyn  1977; Seaquist \etal  1978; Duric \etal 1983;  Hummel \etal
1983; Duric \&  Seaquist 1988; Baan \& Irwin  1995).  This morphology,
which is  also seen in the  X-rays (Fabbiano \etal  1992; Dahlem \etal
Heckman 1998; Pietsch \etal Volger  1998) and in optical line emission
(Ford \etal  1986) has been interpreted  as an outflow  from a compact
central  engine that  interacts  with  the dense  gas  in the  nuclear
environment  (see  discussion in  Filippenko  \&  Sargent 1992).   The
hypothesis  that  this outflow  may  be  driven  by accretion  onto  a
supermassive  black  hole is  reinforced  by  high-resolution VLA  and
global VLBI observations (e.g.,  Irwin \& Seaquist 1988; Trotter \etal
1998;  Sawada-Satoh \etal  2000) that  reveal several  aligned sources
thought to be features of a nuclear jet.

Our EVN  observations detect  a slightly resolved,  $\sim$15~mJy radio
core, in addition to a  weak ($\sim$2~mJy) extension 17~mas toward the
Southeast.   From  its  flux  density  and position  relative  to  the
extended  emission,  we identify  the  core  with the  peaked-spectrum
component  ``B'' observed  in  the  radio maps  of  Irwin \&  Seaquist
(1988),  Trotter \etal  (1998), and  Sawada-Satoh \etal  (2000).  From
comparison with  the above-mentioned data, the 5\,GHz  flux density of
component B  has been constant,  within the errors, from  1986 through
1997.  On the  other hand, the  8\,GHz values of Trotter  \etal (1998)
and   Sawada-Satoh  \etal  (2000)   suggest  variability   at  shorter
wavelengths.   Based   on  the  probable  variability   and  the  high
brightness temperature measured in this paper, we identify component B
with  the core  of  NGC\,3079.   The weak  extended  emission that  we
observe Southeast of B, but which is not well constrained by our data,
may  be associated  with components  ``A'' and  ``C'' as  seen  in the
previous studies. Based on  their spectra, the multi-component nature,
and the position  of the extended structure, Trotter  \etal (1998) and
Sawada-Satoh \etal  (2000) argue that A  and C can  be identified with
the fading and/or expanding components in a radio jet.

\subsection{NGC\,3998}  

Low-resolution radio  observations reveal arcminute-scale  emission in
NGC\,3998:  a  structure  consisting   of  a  core  and  double  lobes
($\sim$4\arcmin\  or  20~kpc)  was  detected  with an  overall  PA  of
$\sim$0\deg\  to $-$15\deg\  (Hummel  1980; Wrobel  \& Heeschen  1984;
Wrobel 1991),  slightly misaligned relative to the  galaxy minor axis.
The radio core is variable and has a flat spectrum (e.g., Hummel \etal
1984).

Our high-resolution  EVN observations detect  a strong ($\sim$83~mJy),
slightly resolved core, consistent  with the VLBI measurement (86~mJy)
obtained by  Hummel \etal (1982).   The core displays a  weak northern
extension, which  we suspect is  the innermost part of  the postulated
kpc-scale outflow (Hummel 1980).  The presence of a compact, variable,
flat-spectrum, high-brightness temperature  radio core associated with
a jetlike  extension constitutes  compelling evidence for  an AGN-type
source in  this LINER.  This  evidence is further strengthened  by the
detection of an X-ray (Roberts  \& Warwick 2000; Terashima \etal 2000;
Pellegrini  \etal 2000)  and ultraviolet  (Fabbiano \etal  1994) point
source, by  the presence of a  broad H${\alpha}$ emission  line (Ho et
al.   1997c), and  by the  inferred presence  of  a 10$^8$~M$_{\odot}$
black hole (Dressel \etal 2000).

%TABLE 3 - Properties of the Radio Cores
 
\begin{table*}[!ht]
\footnotesize
\begin{center}
\parbox[b]{17.1cm}{\vspace{2mm}
\caption{{\bf Properties of the Radio Cores}.
Column 1: Source name.
Column 2: Monochromatic power of the core at 5 GHz, computed from the peak flux.
Column 3: Spectral luminosity at 5 GHz, $L_{\rm 5\,GHz} \equiv \nu P_{\rm 5\,GHz}$.
Column 4: Radius.
Column 5: Peak brightness temperature.
Column 6: Variability information (see text).
Column 7: Non-simultaneous 2- or 3-point radio spectral index.
Column 8: References used to calculate $\alpha_{\rm r}$:
(1) Slee \etal  1994 (2.3 and  8.4\,GHz);
(2) Trotter \etal 1998 (5, 8 and 22\,GHz);
(3) Wrobel \& Heeschen 1984 (1.5, 4.9 and 15\,GHz);
(4) Nagar \etal 2000 and this paper (5 and 15\,GHz).
Column 9: The logarithmic ratio of the flux densities at 60 $\mu$m and 1.4\,GHz.
Column 10: The infrared spectral index between 25 and 60 $\mu$m.
}
}
\begin{tabular}{cccccccccc}
\hline
Source    & $P_{\rm 5\,GHz}$ & $L_{\rm 5\,GHz}$  & $r$      & $T_{\rm B}$      & Variable & $\alpha_{\rm r}$ & Ref. & $u$  & $\alpha_{\rm IR}$ \\
          & (W Hz$^{-1}$)    & (erg s$^{-1}$)    & (lty $\times$ lty)   & (K)              &          &                  &      &      &                   \\
(1)       & (2)              & (3)                  & (4)              & (5)      & (6)              & (7)  & (8)  & (9) & (10)              \\
\hline
NGC\,2911 & $3.98 \times 10^{21}$ & $1.99 \times 10^{38}$ & $<1.4 \times 0.9$ & $> 1 \times 10^8$& yes\ &  0.21        & 1 & 0.69 & $<$1.11\\
NGC\,3079 & $6.87 \times 10^{20}$ & $3.44 \times 10^{37}$ & $<0.6 \times 0.4$ & $> 1 \times 10^8$& yes? & $-$1.68/0.82 & 2 & 1.81 & \  3.02\\
NGC\,3998 & $4.37 \times 10^{21}$ & $2.19 \times 10^{38}$ & $<0.7 \times 0.4$ & $> 6 \times 10^8$& yes\ & 0.15/0.32    & 3 & 0.75 & \  1.78\\
NGC\,6500 & $1.29 \times 10^{22}$ & $6.45 \times 10^{38}$ & $<1.6 \times 0.6$ & $> 5 \times 10^8$& yes\ & $-$0.03      & 4 & 0.54 & \  2.12\\
\hline
\end{tabular}
\end{center}
\end{table*}
\normalsize

\subsection{NGC\,6500}  

Arcsecond-scale  radio emission has  been detected  in this  source by
Unger  \etal  (1989),  who  find  two-sided extended  emission  up  to
5\arcsec\ from the nucleus at PA=140\deg, roughly perpendicular to the
galaxy major axis.  Unlike the  highly collimated jet emission seen in
radio galaxies, this arcsecond-scale emission has a wide opening angle
of  $\sim$60\deg.   The  radio  morphology  has  been  interpreted  as
evidence  for an outflow  along the  minor axis  of the  galaxy (Unger
\etal 1989), similar  to that seen in NGC\,3079  (see Sect. 5.2).  On the
other hand,  the central 1\arcsecpoint3 extended emission  seen in the
408\,MHz  and  1.7\,GHz MERLIN  maps  (Unger  \etal  1989) is  roughly
aligned with the major axis of the galaxy, at PA = 55\deg\ and 70\deg,
respectively, consistent  with early  VLBI experiments by  Jones \etal
(1981, 1982).   More recently, the 5\,GHz VLBA  observations of Falcke
\etal (2000) show a core straddled by two-sided emission (overall size
20~mas, PA = 39\deg), aligned to within 9\deg\ of the extended optical
emission-line gas  detected by  Gonzal\'ez Delgado \&  P\'erez (1996).
The  misalignment between the  small and  large scales  may be  due to
projection effects,  or, alternatively, the  jet may be  disrupted and
redirected very near the core.  NGC\,6500 is also known to be variable
in the radio  (e.g., Hummel \etal 1984) and has  a flat radio spectrum
(e.g., Falcke \etal 2000).

Our  high-resolution   observations  show  a   strong  ($\sim$84~mJy),
marginally resolved core consistent  with the VLBA (Falcke \etal 2000)
and VLA (Nagar \etal 2000) measurements. We do not detect the extended
emission reported by  Falcke \etal (2000), most likely  because of the
high  noise level in  our EVN  data.  The  presence of  an unresolved,
variable,   flat-spectrum,  high-brightness  temperature   radio  core
associated with jetlike emission again provide compelling evidence for
an AGN-type source, especially when considered in conjunction with the
X-ray (Barth \etal 1997) and ultraviolet (Barth \etal 1997, 1998; Maoz
\etal 1998) detections.

\section{Discussion and Summary}

The physical  origin of  LINERs has been  a controversial  topic since
Heckman (1980)  identified them as  a major constituent of  the galaxy
population.    In  recent   years,   high-resolution,  multiwavelength
observations  have contributed  greatly to  elucidating the  nature of
these enigmatic objects.  As discussed recently by Barth (2001) and Ho
(2001), there is  now little doubt that ``type~1''  LINERs (those with
detectable broad emission lines) are genuine low-luminosity AGNs.

An  outstanding issue yet  to be  resolved is  the AGN  fraction among
narrow-lined, ``type~2'' LINERs and so-called transition objects.  The
traditional optical  diagnostic emission lines  are largely degenerate
with respect to  a number of the ionization  mechanisms that have been
proposed (see Sect. 1).  While the ultraviolet region can be advantageous
compared to the optical, the detection  rate in this band is low (Maoz
\etal 1995;  Barth \etal  1998), most likely  due to a  combination of
dust extinction (Barth \etal 1998; Pogge \etal 2000) and the intrinsic
weakness of LINERs  in this spectral region (Ho  1999b, 2001).  In the
few cases  where ultraviolet  spectroscopy is available,  the evidence
for AGNs has been mixed (Maoz \etal 1998; Shields \etal 2001).

A  method widely  used to  discriminate AGNs  from starburst-dominated
sources compares the relative strength of the far-infrared flux to the
radio  flux.   For  ``normal''  or star-forming  galaxies,  Condon  \&
Broderick  (1988) find  that the  distribution of  the  $u$ parameter,
defined as  the logarithmic ratio  of the flux densities  at 60~$\mu$m
and  1.4\,GHz, peaks at  $u\,\approx\,2.0$, with  a tail  toward lower
values  of $u$  (excess  radio emission)  due  to galaxies  containing
prominent  AGNs.   AGNs also  generally  have ``warmer''  far-infrared
spectra  compared  to star-forming  galaxies  (e.g.,  de~Grijp \etal
1985), which  are characterized  by $\alpha_{\rm IR}  \approx 2.3-3.0$
between  25 and  60~$\mu$m  (Condon \&  Broderick  1988; Condon  \etal
1991).  We  have calculated $u$  and $\alpha_{\rm IR}$ for  our sample
(Table~3) using  the far-infrared measurements tabulated in  Ho et al.
(1997a) and the integrated 1.4\,GHz  flux densities from the NVSS (see
Table~1).  Indeed,  all three galaxies  with a dominant  nuclear radio
component (NGC\,2911, NGC\,3998, and NGC\,6500) do have a low value of
the  $u$ parameter  ($< 1$)  and a  relatively flat  infrared spectrum
($\alpha_{\rm IR}$ \lax 2).  (NGC~3079  is more ambiguous, but this is
not surprising in view of  the circumstantial evidence for strong star
formation  suggested   by  its  optical   morphology.)   However,  the
application  of these  infrared  diagnostics depends  on the  relative
strength of the nuclear radio emission.

The X-ray band,  especially for energies above 2~keV,  provides a more
promising tool to probe the  central source in LINERs.  However, until
the recent  advent of  {\it Chandra}\ (Ho  \etal 2001),  previous hard
X-ray observations of these  sources (e.g., Ptak \etal 1999; Terashima
\etal 2000) relied on the  coarse beam presented by {\it ASCA}.  Given
the complexity of the X-ray structure in the central regions of nearby
galaxies (e.g.,  Ho \etal 2001),  the low-resolution {\it  ASCA}\ data
also can  yield ambiguous results  for the less prominent  nuclei that
usually characterize type~2 LINERs.

This paper  illustrates that radio  VLBI observations can be  added to
the arsenal of tools to  tackle the LINER problem, and, moreover, that
the radio data {\it  alone}\ can give meaningful physical constraints.
We  have obtained mas-resolution  5~GHz observations  of a  small, but
representative,  sample  of  LINERs.   The  radio maps  enable  us  to
pinpoint highly compact central cores  with sizes \lax 1.5 light year,
which in turn place  stringent lower limits on brightness temperatures
(\gax $10^8$  K) that definitively rule  out a thermal  origin for the
radio emission.  The nonthermal,  AGN-like nature of the radio sources
is further suggested by  other radio characteristics found in previous
observations.  These include the detection of source variability, flat
or  inverted  spectra\footnote{The   spectral  indices  summarized  in
Table~2  derive  from  non-simultaneous  observations,  and  therefore
should  be interpreted  with some  caution  in light  of the  variable
nature  of the  sources.  Ulvestad  \& Ho  (2001)  found flat-spectrum
cores  in  three LINERs  that  were  observed  simultaneously at  four
frequencies  with the  VLBA.}, and  in three  out of  the  four cases,
morphological evidence  for jetlike features or outflows.   All of the
above  are hallmark  features of  ``classical'' AGNs,  the distinction
being  that the  luminosities of  our  sources are  several orders  of
magnitude lower than those observed in traditional radio galaxies.

\section{Acknowledgments}

M.~E.~F.  acknowledges support from the Funda\c c\~ao para a Ci\^encia
e Tecnologia, Minist\'erio da Ci\^encia e Tecnologia, Portugal through
the  grant  PRAXIS XXI/BD/15830/98.   M.~E.~F.   would  like to  thank
Denise  Gabuzda and  Lorant Sjouwerman  at JIVE  (Joint  Institute for
VLBI) for their  help on the data reduction.  We  also thank the staff
of the EVN observatories and the MkIII VLBI Correlator at MPIfR, Bonn.
The work of  L.~C.~H.  is partly funded by NASA  grants from the Space
Telescope  Science  Institute  (operated  by AURA,  Inc.,  under  NASA
contract  NAS5-26555).  This research  has made  extensive use  of NED
(NASA/IPAC  Extragalactic  Database), which  is  operated  by the  Jet
Propulsion  Laboratory,  California   Institute  of  Tecnology,  under
contract with NASA.

{}


\begin{thebibliography}{}

\bibitem[]{}
Alonso-Herrero A., Rieke M.~J., Rieke G.~H., \& Shields, J.~C., 2000, 
ApJ, 530, 688

\bibitem[]{}
Baan W.~A., \& Irwin J.~A., 1995, ApJ, 446, 602

\bibitem[]{}
Barth A.~J., 2001, in Maiolino R., Marconi A., \& Nagar N., ed., 
Issues in Unification of AGNs, San Francisco: ASP, in press

\bibitem[]{}
Barth A.~J., Filippenko A.~V., \& Moran E.~C., 1999, ApJ, 525, 673
 
\bibitem[]{}
Barth A.~J., Ho L.~C., Filippenko A.~V., \& Sargent W.~L.~W., 1998, ApJ, 496, 133

\bibitem[]{}
Barth A.~J., Reichert G.~A., Ho L.~C., Shields J.~C., Filippenko A.~V., 
\& Puchnarewicz E.~M., 1997, AJ, 114, 2313
 
\bibitem[]{}
Barth A.~J., \& Shields J.~C., 2000, PASP, 112, 753

\bibitem[]{}
Becker R.~H., White R.~L., \& Edwards A.~L., 1991, ApJS, 75, 1

\bibitem[]{}
Becker R.~H., White R.~L., \& Helfand D.~J., 1995, ApJ, 450, 559

\bibitem[]{}
Condon J.~J., \& Broderick J.~J., 1988, AJ, 96, 30

\bibitem[]{}
Condon J.~J., Cotton W.~D., Greisen E.~W., Yin Q.~F., Perley R.~A., Taylor 
G.~B., \& Broderick J.~J., 1998, AJ, 115, 1693

\bibitem[]{}
Condon J.~J., Frayer D.~T., \& Broderick J.~J., 1991, AJ, 101, 362

\bibitem[]{}
Dahlem M., Weaver K.~A., \& Heckman T.~M., 1998, ApJS, 118, 401  

\bibitem[]{}
de Bruyn A.~G., 1977, A\&A, 58, 221

\bibitem[]{}
de Grijp M.~H.~K., Miley G.~K., Lub J., \& de Jong T., 1985, Nature, 314, 240

\bibitem[]{}
Dopita M.~A., \& Sutherland R.~S., 1995, ApJ, 455, 468

\bibitem[]{}
Dressel L.~L., Ford H.~C., Kriss G.~A., \& Tsvetanov Z.~I., 2000, BAAS, 196, 21.04 

\bibitem[]{}
Duric N., \& Seaquist E.~R., 1988, ApJ, 326, 574

\bibitem[]{}
Duric N., Seaquist E.~R., Crane P.~C., Bignell R.~C., \& Davis L.~E., 1983, 
ApJ, 273, 11

\bibitem[]{}
Fabbiano G., Fassnacht C., \& Trinchieri G., 1994, ApJ, 434, 67

\bibitem[]{}
Fabbiano G., Kim D.-W., \& Trinchieri G., 1992, ApJS, 80, 531

\bibitem[]{}
Falcke H., Nagar N.~M., Wilson A.~S., \& Ulvestad J.~S., 2000, ApJ, 542, 197

\bibitem[]{}
Filippenko A.~V., \& Sargent W.~L.~W., 1985, ApJS, 57, 503

\bibitem[]{}
Filippenko A.~V., \& Sargent W.~L.~W., 1992, AJ, 103, 28

\bibitem[]{}
Filippenko A.~V., \& Terlevich R., 1992, ApJ, 397, L79

\bibitem[]{}
Ford H.~C., Dahari O., Jacoby G.~H., Crane P.~C., \& Ciardullo R., 1986, 
ApJ, 311, L7

\bibitem[]{}
Fosbury R.~A.~E., Mebold U., Goss W.~M., \& Dopita M.~A., 1978, MNRAS, 183, 549

\bibitem[]{}
Gonz\'alez Delgado R.~M., \& P\'erez E., 1996, MNRAS, 281, 1105

\bibitem[]{}
Halderson E.~L., Moran E.~C., Filippenko A.~V., \& Ho, L.~C., 2001, AJ, 122, 637

\bibitem[]{}
Heckman T.~M., 1980, A\&A, 87, 152

\bibitem[]{}
Ho L.~C., 1999a, ApJ, 510, 631

\bibitem[]{}
Ho L.~C., 1999b, ApJ, 516, 672

\bibitem[]{}
Ho L.~C., 2001, in Green R.~F., Khachikian E.~Ye., \& Sanders D.~B., ed., 
IAU Colloq. 184, AGN Surveys, San Francisco: ASP, in press

\bibitem[]{}
Ho L.~C. et al., 2001, ApJ, 549, L51

\bibitem[]{}
Ho L.~C., Filippenko A.~V., \& Sargent W.~L.~W., 1995, ApJS, 98, 477

\bibitem[H97a]{}
Ho L.~C., Filippenko A.~V., \& Sargent W.~L.~W., 1997a, ApJS, 112, 315

\bibitem[H97b]{}
Ho L.~C., Filippenko A.~V., \& Sargent W.~L.~W., 1997b, ApJ, 487, 568

\bibitem[H97]{}
Ho L.~C., Filippenko A.~V., \& Sargent W.~L.~W., Peng C.~Y., 1997c, ApJS, 
112, 391

\bibitem[]{}
Hummel E., 1980, A\&AS, 41, 151

\bibitem[]{}
Hummel E., Fanti C., Parma P., \& Schilizzi R.~T., 1982, A\&A, 114, 400

\bibitem[]{}
Hummel E., van der Hulst J.~M., \& Dickey J.~M., 1984, A\&A, 134, 207

\bibitem[]{}
Hummel E., van Gorkom J.~H., \& Kotanyi C.~G., 1983, ApJ, 267, L5

\bibitem[]{}
Irwin J.~A., \& Seaquist E.~R., 1988, ApJ, 335, 658 

\bibitem[]{}
Jones D.~L., Sramek R.~A., \& Terzian Y., 1982, ApJ, 261, 422

\bibitem[]{}
Jones D.~L., Terzian Y., \& Sramek R.~A., 1981, ApJ, 246, 28

\bibitem[]{}
Maoz D., Filippenko A.~V., Ho L.~C., Rix H.-W., Bahcall J.~N., 
Schneider D.~P., \& Macchetto F.~D., 1995, ApJ, 440, 91

\bibitem[]{}
Maoz D., Koratkar A.~P., Shields J.~C., Ho L.~C., Filippenko A.~V., 
\& Sternberg A., 1998, AJ, 116, 55

\bibitem[]{}
Nagar N.~M., Falcke H., Wilson A.~S., \& Ho L.~C., 2000, ApJ, 542, 186

\bibitem[]{}
Pellegrini S., Cappi M., Bassani L., Della Ceca R., \& Palumbo
G.~G.~C., 2000, A\&A, 360, 878

\bibitem[]{}
Pietsch W., Trinchieri G., \& Volger A., 1998, A\&A, 340, 351

\bibitem[]{}
Pogge R.~W., Maoz D., Ho L.~C., \& Eracleous M., 2000, ApJ, 532, 323

\bibitem[]{}
Ptak A., Serlemitsos P., Yaqoob T., \& Mushotzky R., 1999, ApJS, 120, 179

\bibitem[]{}
Roberts T.~P., \& Warwick R.~S., 2000, MNRAS, 315, 98

\bibitem[]{}
Sadler E.~M., Jenkins C.~R., \& Kotanyi C.~G., 1989, MNRAS, 240, 591

\bibitem[]{}
Sandage A., \& Bedke J., 1994, The Carnegie Atlas of Galaxies 
(Washington, DC: Carnegie Inst. of Washington)

\bibitem[]{}
Sawada-Satoh S., Inoue M., Shibata K.~M., Kameno S., Migenes V., Nakai N., 
\& Diamond P., 2000, PASJ, 52, 421 

\bibitem[]{}
Schilizzi R.~T., Fanti C., Fanti R., \& Parma P., 1983, A\&A, 126, 412

\bibitem[]{}
Seaquist E.~R., Davis L., \& Bignell R.~C., 1978, A\&A, 63, 199

\bibitem[]{}
Shields J.~C., 1992, ApJ, 399, L27

\bibitem[]{}
Shields J.~C., Sabra B.~M., Ho L.~C., Barth A.~J., Filippenko A.~V., 2001, 
in Crenshaw D.~M., Kraemer S.~B., \& George I.~M., ed., Mass Outflow in Active 
Galactic Nuclei: New Perspectives, San Francisco: ASP, in press

\bibitem[]{}
Slee O.~B., Sadler E.~M., Reynolds J.~E., \& Ekers R.~D., 1994, MNRAS, 269, 928

\bibitem[]{}
Terashima Y., Ho L.~C., \& Ptak A.~F., 2000, ApJ, 539, 161 

\bibitem[]{}
Terlevich R., \& Melnick J., 1985, MNRAS, 213, 841

\bibitem[]{}
Terlevich R., Tenorio-Tagle G., Franco J., \& Melnick J., 1992, MNRAS, 255, 71

\bibitem[]{}
Trotter A.~S., Greenhill L.~J., Moran J.~M., Reid M.~J., Irwin J.~A., 
\& Lo K.-Y., 1998, ApJ, 495, 740 

\bibitem[]{}
Tully, R.~B. 1988, Nearby Galaxies Catalog (Cambridge: Cambridge Univ. Press)

\bibitem[]{}
Ulvestad J.~S., \& Ho L.~C., 2001, ApJ, submitted

\bibitem[]{}
Unger S.~W., Pedlar A., \& Hummel E., 1989, A\&A, 208, 14

\bibitem[]{}
van Breugel W.~J.~M., Schilizzi R.~T., Hummel E., \& Kapahi V.~K., 1981, 
A\&A, 96, 310


\bibitem[]{}  van Moorsel  G., Kemball  A.,  \& Greisen  E., 1996,  in
Jacoby G.  H. \& Barnes  J., ed., Astronomical Data  Anlaysis Software
and Systems V, San Francisco: ASP, p.37

\bibitem[]{}
White R.~L., \& Becker R.~H., 1992, ApJS, 79, 331

\bibitem[]{}
Wrobel J.~M., 1991, AJ, 101, 127

\bibitem[]{}
Wrobel J.~M., \& Heeschen D.~S., 1984, ApJ, 287, 41

\bibitem[]{}
Wrobel J.~M., \& Heeschen D.~S., 1991, AJ, 101, 148

\end{thebibliography}
\end{document}